# Probing the atomically diffuse interfaces in core-shell nanoparticles in three dimensions

Zezhou Li[1], Zhiheng Xie[1], Yao Zhang[1], Xilong Mu[1], Hai-jing Yin[1], Ya-wen Zhang[1], Colin Ophus[2], Jihan Zhou[1]

*[1]Beijing National Laboratory for Molecular Sciences, College of Chemistry and Molecular Engineering, Peking University, Beijing, 100871, China.*

*[2]National Center for Electron Microscopy, Molecular Foundry, Lawrence Berkeley National Laboratory, Berkeley, CA 94720, USA.*

*Correspondence and requests for materials should be addressed to J. Z. (email: jhzhou@pku.edu.cn)*

**Deciphering the three-dimensional atomic structure of solid-solid interfaces in core-shell nanomaterials is the key to understand their remarkable catalytical, optical and electronic properties. Here, we probe the three-dimensional atomic structures of palladium-platinum core-shell nanoparticles at the single-atom level using atomic resolution electron tomography. We successfully quantify the rich structural variety of core-shell nanoparticles including bond length, coordination number, local bond orientation order, grain boundary, and five-fold symmetry, all in 3D at atomic resolution. Instead of forming an atomically-sharp boundary, the core-shell interface is atomically diffuse with an average thickness of 4.2 Å, irrespective of the particle's size, morphology, or crystallographic texture. The high concentration of Pd in the interface is highly related to the electric double**

**layers of the Pd seeds. These results advance our understanding of core-shell structures at the fundamental level, providing potential strategies into nanomaterial manipulation and chemical property regulation.**

## Introduction

Core-shell nanoparticles (CS-NPs) are attracting significant interest in chemistry and material science, and are widely used in many applications such as catalysis, optics, electronics, and biomedical applications due to their novel optical, magnetic and electronic characteristics[1-6]. The unique and versatile functionalities of CS-NPs arise from the physical and chemical properties of both the core and the shell[7,8], determined by the three-dimensional (3D) arrangement of the atoms, particularly at the core-shell interfaces. The core and shell structures can be modified by tailoring the composition, size and morphology of each accurately, leading to a large number of combinatorial properties[9,10]. For example, by managing the surface strain[11] or designing the thickness[12] of the shell, the catalytic performance of bimetallic CS-NPs can be tuned in a controlled manner.

The characterization of CS-NPs is critical to understand structure-functionality correlations, which in turn influence the design of materials with improved properties. In the past decades, through rapid development of electron microscopy, both structural and elemental information of CS-NPs can be routinely characterized at atomic resolution using either transmission electron microscopy (TEM) or scanning transmission electron microscopy (STEM)[13-18]. However, investigation of nanostructures from two-dimensional images could give deceptive structural information due to the overlap of critical features[19]. Thus, crucial questions about core-shell interface structures still exist[20,21]. The 3D atomic structure of most CS-NPs and the atomic arrangement of the solid-solid interfaces between the core and the shell remain unknown, and it is still unclear how this interface affects the surface properties. It has been reported that the geometric misfit introduces interfacial strain during heteroepitaxial growth of the shell in solution to form multigrain nanocrystals[22]. It is also important to show in 3D how the deformation of the lattice maintains coherency across the interfaces in the core-shell heteroepitaxial structures[23-25].

To address these critical issues of core-shell interfacial structure, here we synthesized polycrystalline pentagonal bipyramid and monocrystalline truncated octahedron palladium-platinum CS-NPs using two-step chemical reduction, and then probed the 3D atomic structures of CS-NPs at the single-atom level. The 3D atomic coordinates and chemical composition of Pd@Pt CS-NPs with different morphology and size were determined using atomic resolution electron tomography (AET), which has been a powerful tool for atomic-scale structural characterization in 3D and even 4D[19,26-29]. We discovered a 3D atomically diffuse core-shell interface. We found that the average thickness of this diffuse interface in Pd@Pt CS-NPs is 4.2 Å, irrespective of the particle's shape, size, or crystallographic texture. The radially-averaged concentration of Pd in the interface drops in a two-step manner which is correlated to the electric double layers. The rich structural information including both the five-fold symmetry and monocrystalline CS-NPs confirmed the heteroepitaxial growth of Pt shell on the Pd seeds. The 3D strain tensor analysis shows no significant difference between the interface and the whole particle. Atoms in the five-fold co-axis form pentagonal bipyramids with their nearest neighbors, which penetrate through the core-shell interface with an average distorted angle smaller than 5 degrees. We discuss the origin of the diffuse interfaces by correlating these observations with possible growth mechanisms. These findings pave the way for further studies of nanomaterial manipulation and chemical property regulation.

**Results**

**Determining the 3D atomic coordinates/species of core-shell nanoparticles**

Pd@Pt CS-NPs with different morphologies were synthesized by a two-step chemical reduction method[30] (Supplementary Fig. 1, Methods). Energy-dispersive X-ray (EDX) spectroscopy mappings show well-defined Pd@Pt core-shell structures with no oxygen signals on the surface (Supplementary Fig. 1). Tomographic tilt series (Supplementary Figs. 2-4) were acquired from several nanoparticles with different morphologies using an aberration-corrected transmission electron microscope in annular dark-field (ADF) STEM mode (Supplementary Table 2). Negligible structure change was observed throughout the tilting experiment (Supplementary Fig. 5), showing the CS-NPs were stable during the tomographic tilting imaging experiment under the total electron dose. We focus on three representative particles with different shapes; namely, pentagonal bipyramid (PB), elongated pentagonal bipyramid (EPB) and truncated octahedron (TO) in this study.

After pre-processing and image denoising, 3D reconstructions were computed from the tilt series using an iterative algorithm described elsewhere[28]. All the 3D atom coordinates were determined and classified (Methods, Supplementary Table 2). PB, EPB and TO particles are different in size, with a large variety in both core and shell atom numbers. Figs. 1a,b and Supplementary movie 1 show the experimental 3D atom models of PB, EPB and TO particles with Pd atoms in blue and Pt atoms in yellow. The insets in Fig. 1a show three corresponding Johnson solids representing the shape of each particle. Despite of the variety in size and shape, the interfaces between the Pd core and the Pt shell show diffuse chemical profiles, without a defined sharp boundary in chemical composition for all three particles (Fig. 1c, Supplementary movie 2). The diffuse core-shell interfaces with single Pd atoms spreading into the Pt shell were observed in the internal atomic slices in all three particles (Fig. 1c). From the experimental 3D coordinates, we quantitatively characterized the coordination number (CN), bond length and local bond orientation order (BOO) parameters at the single-atom level (Figs. 1d-f). The CN of surface atoms at {111} facets are almost all 9, while CN values smaller than 8 are common at the edges and corners. Almost no difference was observed in the distributions of the Pd-Pd, Pt-Pt and total bond lengths (Supplementary Figs. 6a-c). The mean bond lengths of three particles are 2.76, 2.76 and 2.77 Å for PB, EPB and TO, respectively, close to the standard Pt-Pt bond length of pure Pt metal (2.774 Å)[31]. This is considerable because in all three particles Pt atoms are a few thousands numbers more than Pd atoms. We also observed a small lattice compression of the Pt shell (Supplementary Figs. 6d-f) which is attributed to the effect of surface tension and strong bonding of surface atoms[32]. We calculated the BOO maps using polyhedral template matching[33] (Methods). The TO particle shows a fcc single crystal structure, whereas the pentagonal symmetric PB and EPB particles have five twined fcc grains with hcp-like boundaries (Fig. 1f).

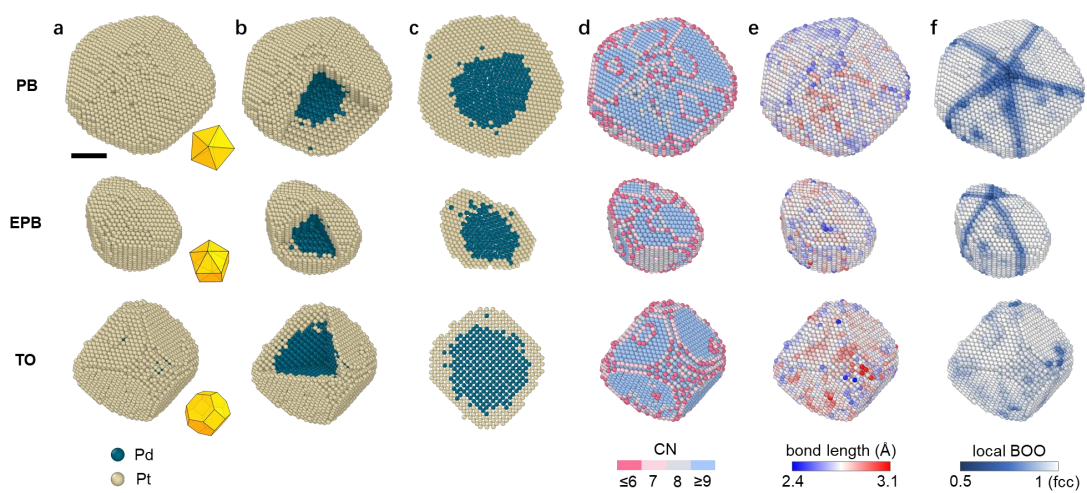

**Fig. 1 | 3D atomic structures and chemical compositions of three core-shell nanoparticles with different morphologies. a-f**, 3D surface morphology (**a**), core-shell structure (**b**), internal atomic slice (**c**), CN (**d**), bond length (**e**) and local BOO parameters (**f**) of PB, EPB and TO particles were identified from the coordinates determined by AET at the single-atom level. Inserts in **a** are the geometric Johnson solid models of three particles. For the structural order, BOO = 1 corresponds to a perfect fcc lattice. Scale bar, 2 nm.

**3D atomically diffuse core-shell interfaces**

To better visualize the internal 3D core-shell interface, we divided the atomic coordinates of three particles into slices four atom-layer-thick (Figs. 2a-c). The 3D interfaces between Pd and Pt are diffuse with a consistent thickness, and in addition there are isolated Pd atoms scattered in the Pt shell. Fig. 2d and Supplementary Figs. 7a-b show the similar distributions of isolated Pd atoms in the TO, PB and EPB particles. However, very few isolated Pt atoms are found in the Pd core. One representative cut-out of the atomic model of TO particle is shown in Fig. 2e. To quantitatively confirm the thickness of the diffuse core-shell interface, we calculated the concentration of chemical composition and mean CN at a depth step of 1.03 Å. For convenience, we defined zero depth at the isosurface where the Pd to Pt ratio equals one (Supplementary Fig. 7c). The radially-averaged concentration of Pd in all three particles drops quickly from the Pd core to the center of the interface, and then decreases at a much slower rate from the center of the interface into the Pt outer shell (Fig. 2f). The interface area was mainly distributed in the Pt shell, with many Pd atoms dispersed into Pt shell, which is shown in the atomic internal slices from all experimental models in Figs. 2a-c. We measured the thickness of the CS interface using Pd concentration between 5% and 95% as the defined thickness range, since AET has been proven to achieve accurate chemical species information, with 95% consistency[27]. The thicknesses of CS interfaces in all three NPs are equal to 4.2 Å, irrespective of the morphology (Fig. 2f), roughly three atomic layers thick. To probe the Pt and Pd coordination, we analyzed the number of nearest-neighbor Pd around a Pt atom and the number of nearest neighbor Pt around a Pd atom in each layer, termed CN-Pt$_{Pd}$ and CN-Pd$_{Pt}$, respectively. CN-Pd$_{Pt}$ = 12 means a Pd site is surrounded by 12 Pt atoms, indicating that this Pd atom is isolated in a pure Pt lattice. Fig. 2g shows the mean CN-Pt$_{Pd}$ decreases while the mean CN-Pd$_{Pt}$ increases from core to shell, indicating that Pd and Pt atoms are diffusively mixed in the interfaces, rather than forming an atomically-sharp boundary. We counted all the isolated atoms and clusters in three particles (Fig. 2h); and 82.9% of them are single Pd atoms distributed in a 10-Å-thick layer in the Pt shell (Fig. 2i).

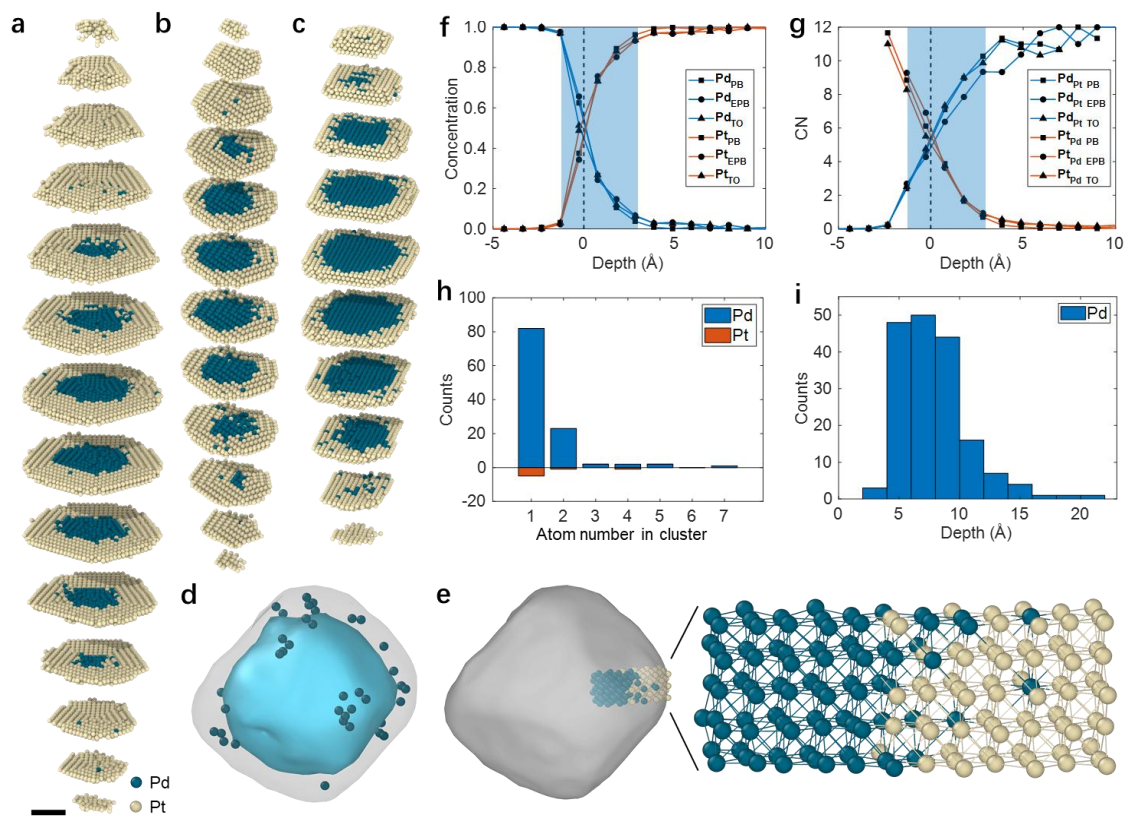

**Fig. 2 | 3D atomically diffuse core-shell interfaces in three particles. a-c**, Atomic

slices of PB (**a**), EPB (**b**) and TO (**c**) show diffuse core-shell interface in each particle. Scale bar for **a-c**, 2nm. **d**, Distribution of isolated Pd atoms in TO particle. The gray and cyan contours show the full nanoparticle and the core, respectively. **e**, A representative cut-out from the Pd-Pt interface in TO shows isolated Pd atoms unconnected to the Pd core. **f**, Radially-averaged Pd and Pt concentrations along the core to the shell, where the diffuse interface is highlighted in blue. The dotted line marks the zero-depth position. **g**, Mean Pd CN around Pt (labeled as CN-Pt$_{Pd}$) and Pt CN around Pd (labeled as CN-Pd$_{Pt}$) in each layer. **h**, Histogram of diffuse atoms at interfaces. **i**, Distribution of the distances from the isolated Pd atoms to the middle of the interface for all particles. Zero depths in **f-i** are defined at the isosurface where the Pd concentration was ~50 % in each particle and marked with dotted black lines (**f, g**).

**Rich structural information and heteroepitaxial growth of Pt shell**

To probe the rich structural information both in the interfaces and the whole particles at atomic resolution, we analyzed the point defects, grain boundaries and edges/corners in all three particles. No vacancies were found in any of the three particles. Figs. 3a-c show the 3D reconstruction volumes of PB, EPB and TO particles projected from the <110>, <110> and <100> direction, respectively. The five grains in PB are almost the same in size (Fig. 3a), while the five grains in the EPB have a large variation in size (Fig. 3b), which is due to the small immature Pd core. We observed alternating {100} and {111} crystal planes with adjacent step edges connecting two neighboring facets and corners (Figs. 3a-c, Supplementary Figs. 8a-c). Instead of growing a sharp corner at each edge, the ridges and edges in all particles form concave grooves to minimize the surface energy (Supplementary Figs. 8d-f).

Figs. 3d-f show the central 2.7-Å-thick reconstruction slices of three reconstructions of PB, EPB and TO, respectively. Each facet on the surface of the particles (solid lines) corresponds to a smaller facet on the surface of Pd core (dotted lines). The morphology of the CS-NPs depends on the morphology of the Pd seeds (Pd core). Figs. 3g-i and Supplementary movie 3 show the cut-outs and corresponding atomic slices of shells in three particles in the direction of <100> (yellow arrows) and <111> (blue arrows) orientations. All our observations suggest the heteroepitaxial growth of the Pt shell on the Pd core, where the Pd and Pt lattice constants are close enough to maintain coherency and make the heteroepitaxial growth process energetically more favorable[21,34].

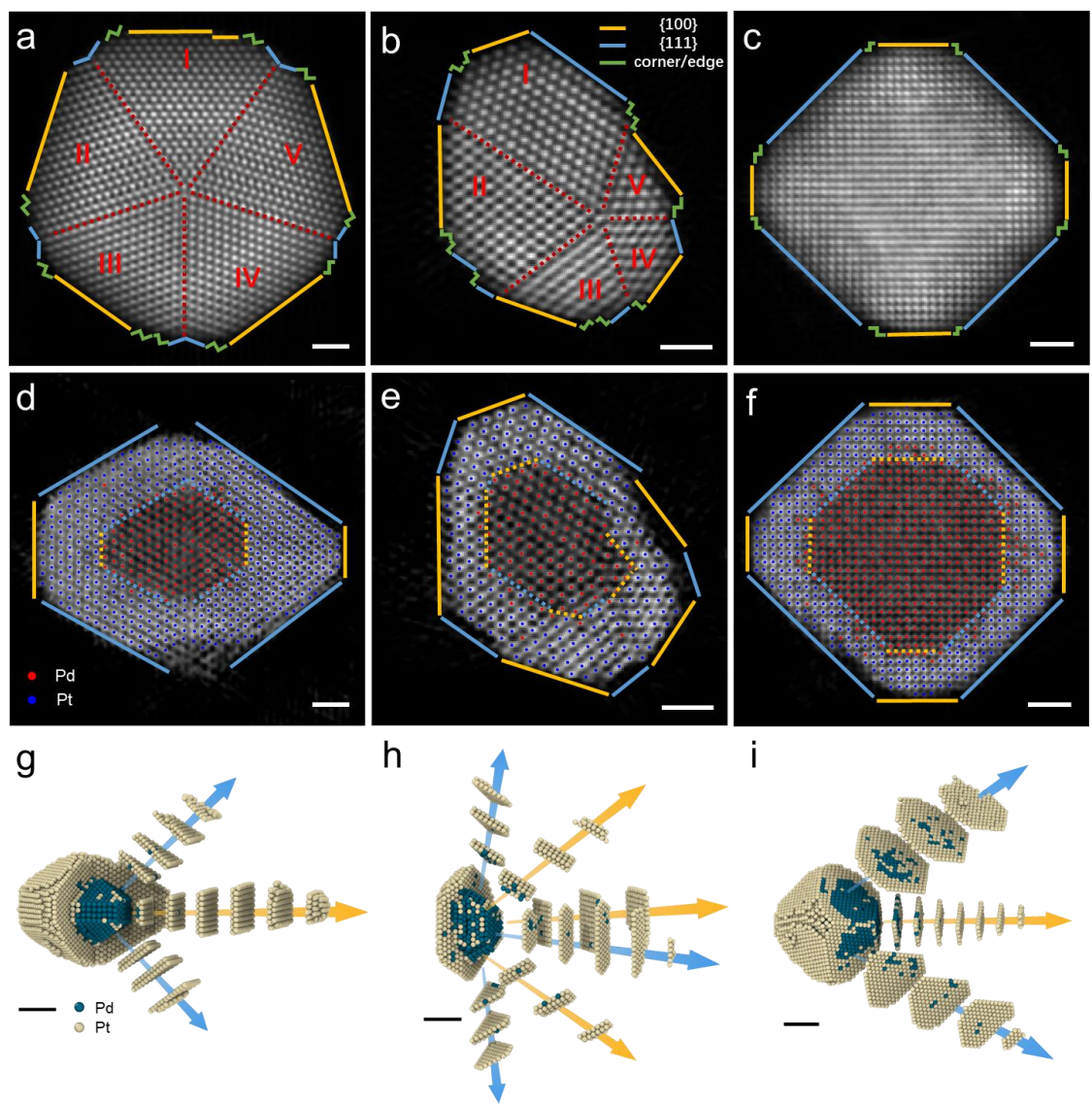

**Fig. 3 | Rich structural information and heteroepitaxial growth of Pt shell in core-shell nanoparticles. a-c**, Projections of 3D reconstruction of PB (**a**), EPB (**b**) and TO (**c**), viewed from <110>, <110> and <100> zone axis, respectively. For PB and EPB, <110> direction is the co-axis of five single crystal grains (five-fold co-axis) direction. Yellow and blue lines show {100} and {111} facets parallel to the view direction. Green fold lines represent corners and steps. Red dashed lines in **a**, **b** highlight the five-fold twin boundaries, and each grain is labelled from I to V. **d-f**, The central 2.7-Å-thick

reconstruction slices of three reconstructions of PB (**d**), EPB (**e**) and TO (**f**), respectively. The red and blue dots show the coordinates of Pd and Pt atoms, respectively. Scale bar in **a-f**, 1 nm. **g-i**, Cut-out atom models of PB (**g**), EPB (**h**) and TO (**i**) with some grains of shell sliced in atomic layers, indicating heteroepitaxial growth of {100} and {111} facets from core to shell. Yellow and blue arrows show the direction of <100> and <111> crystal orientations, respectively. Scale bar in **g-i**, 2 nm.

## 3D strain analysis

In all particles we investigated, the mean bond length distributions at the interfaces are similar to those of the whole particles (Supplementary Fig. 6). To compare the local strains between the interfaces and the whole particles, we calculated the full 3D strain tensors of each nanoparticle. Using the ideal fcc structure with a best-fit lattice[35], we calculated the atomic displacement field, convolved it with a 3.89-Å-wide Gaussian kernel, and then differentiated it to obtain the 3D strain tensors (Supplementary Figs. 9-11). The full 3D strain tensors of three particles are mapped at each atomic position for the central slices all along (110) crystal plane (Figs. 4a, c, e). The strain maps also show that the shell mostly has a larger strain compared to the core. We show all six components of the full strain tensor of PB (Fig. 4b), EPB (Fig. 4d) and TO (Fig. 4f), most of them are around 1%, smaller than most of the strain measurement results reported[24,36-38]. The standard deviations of principal strains ($\varepsilon_{xx}$, $\varepsilon_{yy}$, $\varepsilon_{zz}$) are 0.67%, 0.69%, 0.67% for total PB particle, 0.65%, 0.60%, 0.30% for total EPB particle and

0.33%, 0.55%, 0.98% for total TO particle.

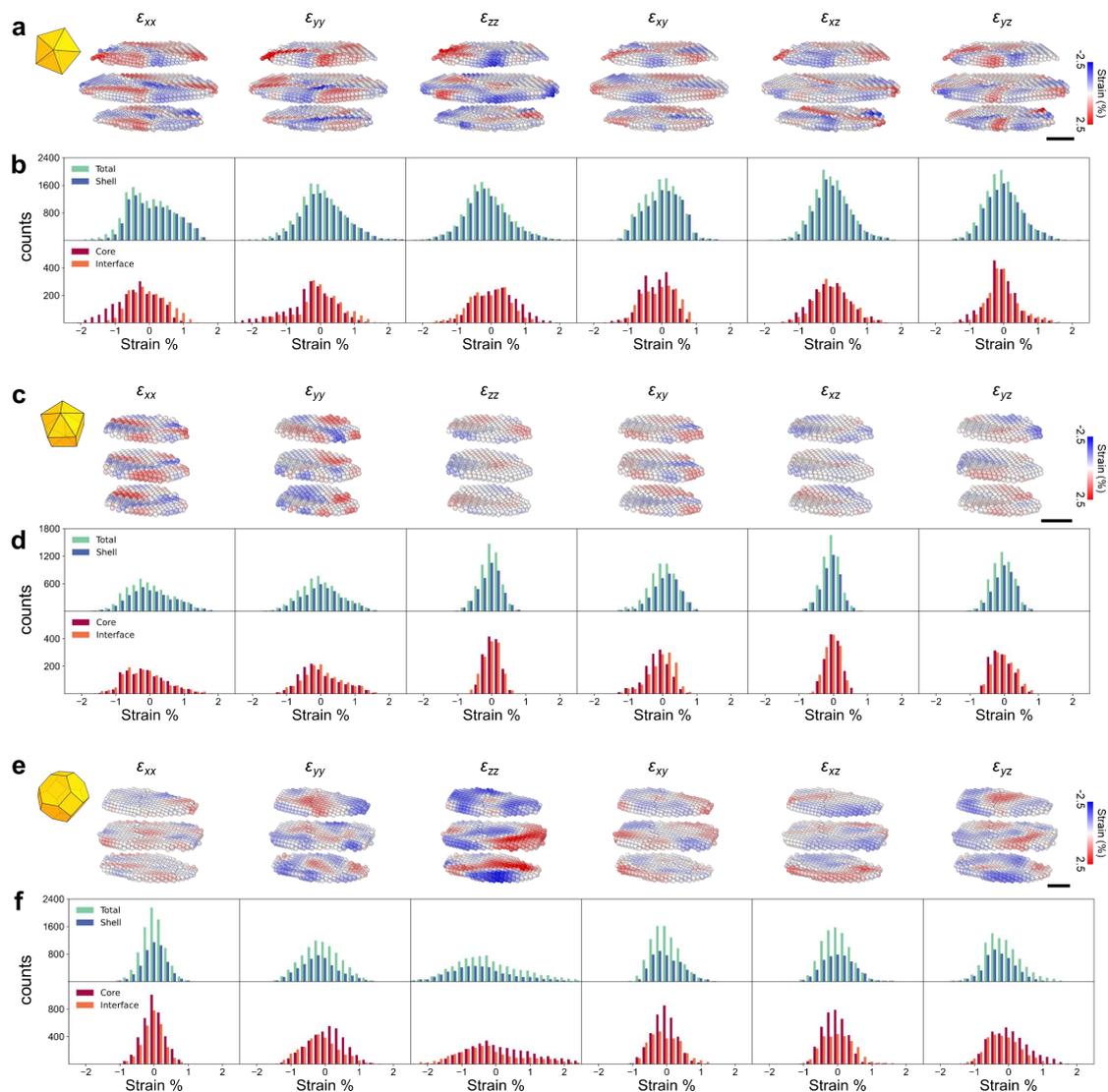

**Fig. 4 | 3D strain tensor analysis of three particles. a**, Sliced maps of the six components of the full strain tensor for PB particle. **b**, Statistics of the strain tensors for total and shell atoms (top), core and core-shell interface atoms (bottom) of PB particle. **c**, Sliced maps of the six components of the full strain tensor for EPB particle. **d**, Statistics of the strain tensors for total and shell atoms (top), core and core-shell interface atoms (bottom) of EPB particle. **e**, Sliced maps of the six components of the full strain tensor

for TO particle. **f**, Statistics of the strain tensors for total and shell atoms (top), core and core-shell interface atoms (bottom) of TO particle. Atomic slices in **a** and **c** are perpendicular to the five-fold co-axis along [110] direction. Slices in **e** are also along (110) crystal plane. In **a**, **c** and **e**, strain is indicated by the color scale, and the scale bar is 2nm.

**Five-fold symmetry structure in EPB and PB nanoparticles**

The five-fold symmetry in EPB and PB particles exists in both the core and shell. We further classified the local structural environments in both particles with polyhedral template matching (Methods). We found five fcc grains in each particle, each sharing an hcp-like twin boundary with its two neighboring grains, and one five-fold co-axis (Figs. 5a, b). The five-fold grain boundary consisting of five {111} crystal planes penetrate throughout the Pd core and Pt shell without significant distortion (Figs. 5c, d). The angles between twined fcc grains are measured as 74.1°, 70.8°, 71.8°, 72.3°, 71.0° for PB particle and 70.6°, 72.2°, 72.7°, 72.1°, 72.4° for EPB particle, which slightly deviated from perfect five-fold symmetry (72°)[39]. The angle mismatch between our measured angles and perfect <110> plain orientation (70.5°) is compensated by the slightly twisted co-axis and grain boundaries. The little lattice mismatch between Pd (1.37 Å) and Pt (1.39 Å) leads to the minor lattice distortion. The atoms in the five-fold co-axis of grain boundaries exhibit a unique local environment with twelve hcp-ordered nearest neighbors. They share ten adjacent boundary atoms to form two pentagonal bipyramids (Fig. 5e). 3D strain tensors analysis shows some significant strain differences between two adjacent grains along this co-axis (Figs. 4a, c), alleviating the distortion in other parts of the particle. We quantified the three bond lengths, from the capping atom bond α, the capping-ring atom bond β and the ring atom bond γ (Fig. 5f). Compared with the ideal pentagonal bipyramid, we found most of the experimental pentagonal bipyramids are compressed in the co-axis direction (α/γ ratios are 0.99 for PB and 0.98 for EPB, smaller than standard 1.05). The β/γ ratios are both 0.99 for PB and EPB, showing almost ideal capping pyramids for the whole five-fold axis (Fig. 5g). We also measured the angle ($\theta$) between the α bond and the plane normal vector of five ring atoms, where the mean $\theta$ is 5.2° for PB and 3.9° for EPB, indicating that the pentagonal bipyramids in the five-fold co-axis are distorted (Fig. 5h).

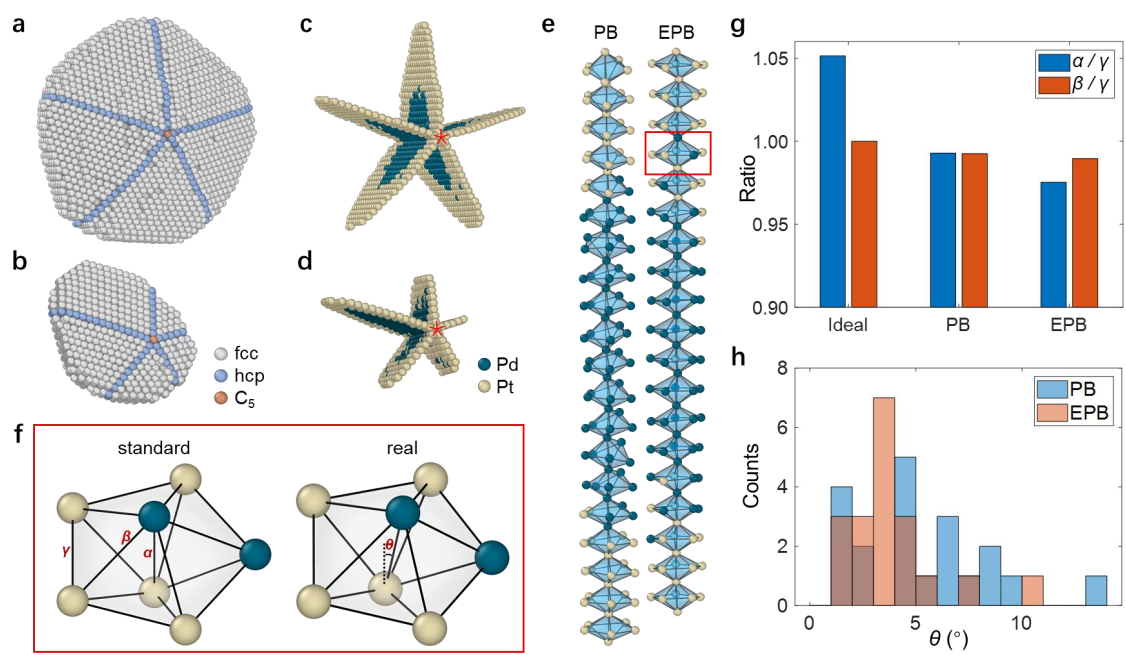

**Fig. 5 | Five-fold symmetry structure in EPB and PB nanoparticles. a-b**, The local

16...

structures in the PB (**a**) and EPB (**b**) NPs show five fcc grains and a continuous hcp twin boundary bordering each pair of grains. **c-d**, The five-fold twin boundary and the axis in the middle passes through both core and shell of PB (**c**) and EPB (**d**), respectively. **e**, Five-fold co-axis of EPB and its nearest neighboring coordination polyhedrons. All atoms in the five-fold co-axis are twelve-coordinated in a pentagonal bipyramid structure. **f**, An experimental pentagonal bipyramid (right, enlarged from red boxed in **e**) and corresponding ideal pentagonal bipyramid Johnson solid (left). The α, β and γ represent the capping, capping-ring and ring atom bonds, respectively. The α bond and the plane of five ring atoms form an angle ($\theta$) in real pentagonal bipyramids. **g**, The α/γ and β/γ ratios of an ideal pentagonal bipyramid, PB and EPB. **h**, Average $\theta$ of the five-fold axis in PB and EPB.

## Discussion and outlook

Although TEM/STEM has been employed to explore the growth mechanism of CS structures in 2D[10,18,40-42], it was speculated that the intermixing of core and shell atoms could happen before a perfect shell was formed[43]. Our experimental study of 3D interfaces in CS-NPs reveals several observations at the single-atom level. The CS interfaces are diffuse in different NPs with almost the same thickness. Isolated Pd atoms in the Pt shell are far more common than vice versa. The radially-averaged concentration of Pd drops very fast between -1.2 Å and 0.8 Å depth away from the core and eventually decreases in a much slower rate after 0.8 Å (Fig. 2f, Supplementary Fig.

7c). The striking similarities among the core-shell interfaces of three particles, irrespective of the size, morphology, or crystallographic texture of the particles, suggest a potential mechanism during the chemical reduction in wet chemistry to form the diffuse interfaces.

We have examined the influence of heating procedures during sample preparation. We baked the CS-NPs at 180 °C in vacuum for much longer times, up to 48 hrs. The 3D atomic models from two independent measurements of the same nanoparticle before and after baking were obtained using the same procedures. The same atomic slice in both models shows that the diffuse Pd atoms in the Pt shell remain the same position and chemical environment (Supplementary Fig. 12, the yellow circles). No nucleation or phase transition was observed during heating. It is been reported that the core-shell structure remains at 400 °C[44]. Thus, the diffuse Pd in the interface does not originate from the atomic motion due to the heating during the synthesis or baking process.

It is notable that we washed the Pd seeds and re-dispersed them in fresh oleylamine to get rid of any extra Pd(II) salts for the shell growth. The size distributions between Pd seeds and the final Pd cores in CS-NPs (Supplementary Fig. 1f) are almost identical, indicating that the etching of Pd with Pt(IV) was negligible. The diffuse Pd atoms in the interface cannot solely come from the galvanic replacement of the Pd seeds with Pt(IV) precursors under a reducing environment, as this would result in a smaller Pd core[45,46]. The local dissolved Pd(II) ions in the electrochemical layers surrounding Pd seeds must participate in the growth of the interface. To further corroborate our observations, we estimated the Debye length $\kappa^{-1}$ of the shell growth system in oleylamine by the following equation[47]:

$$\kappa^{-1} = \sqrt{\frac{\varepsilon_r \varepsilon_0 k_B T}{2 N_A I e^2}}$$

Where $\varepsilon_r$ is the relative permittivity of oleylamine, $\varepsilon_0$ is the vacuum permittivity, $k_B$ is the Boltzmann constant, $T$ is the absolute temperature, $N_A$ is the Avogadro constant, $I$ is the ionic strength and e is the electron charge. The $\varepsilon_r$ of oleylamine is ~3.1 according to literature[48], and the Debye length $\kappa^{-1}$ approximately equals to 10 Å. The Debye length represents the thickness of the electric double layer, which is the same order of magnitude with the thickness of the diffuse interfaces measured in three particles (4.2 Å). We hypothesized that the electric double layer could serve as a reservoir of Pd(II) to be co-reduced during the growth of Pt shell. The high concentration of Pd(II) ions near the interface is highly related to the electric double layers of the Pd seeds. This can also explain the origin of a similar interface in the synthesis of PB and EPB particles, where a less oxidative Pt(acac)$_2$ is used as the precursor of the Pt shells. We show a schematic diagram of the possible formation mechanism of the diffuse interface (Fig. 6).

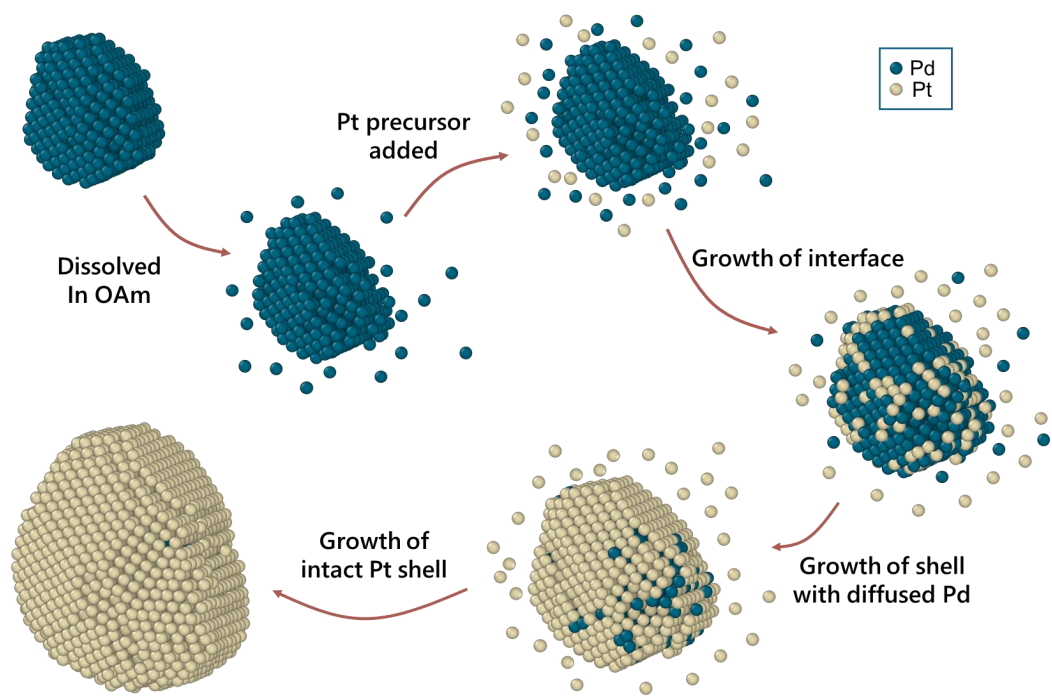

**Fig. 6 | Schematic of the proposed mechanism of core-shell growth for Pd@Pt CS-NPs.** The Pd seed was dispersed in oleylamine. A few surface Pd atoms was ionized and formed an electric double layer around the seed. After the Pt precursor was added to the solution, Pt and Pd ions were reduced simultaneously on the surface of the Pd seed to form the diffuse Pd-Pt interface with some possible substitution of Pd(0) to Pd(II) by Pt(IV). After the core was completely wrapped by Pt atoms, the shell continued to grow heteroepitaxially on the core to form the final particle.

In conclusion, we quantified rich structural variety of Pd@Pt core-shell nanoparticles including bond length, coordination number, local bond orientation order, grain boundary, and five-fold symmetry, all in 3D at atomic resolution. Our results confirmed the heteroepitaxial growth of the core-shell structure. We quantitatively showed the unique atomically diffuse interface in CS-NPs synthesized with wet chemistry with unprecedented 3D detail. Our results suggest that the diffuse interface is highly related to the electric double layers of the seeds, which could possibly be tuned by manipulating the Debye length. The diffuse layer could influence the electronic structures of the shell, and further affect the applications of core-shell structures in catalysis, optics and electronics[20,49-52]. These results will advance our understanding of structure-property relationships of core-shell structures at the fundamental level.

**Acknowledgements** This work was supported by the National Natural Science Foundation of China (Grant No. 22172003 & 21832001) and High-performance Computing Platform of Peking University. We thank the Electron Microscopy Laboratory at Peking University for the use of the aberration-corrected electron microscope. H. Y. and Y. Z. thank the support of National Key R&D Program of China (No. 2021YFA1501100) and Beijing National Laboratory for Molecular Sciences (BNLMS-CXXM-202104). Work at the Molecular Foundry was supported by the Office of Science, Office of Basic Energy Sciences, of the U.S. Department of Energy under Contract No. DE-AC02-05CH11231.

**Author contributions** J. Z. conceived the idea and directed the study. Z. L. performed the AET experiments. X. M., Z. L., H.-J. Y. and Y.-W. Z. and J. Z. discussed/synthesized the Pd@Pt core-shell nanoparticles. Z. L., Z. X., Y. Z. and J. Z. discussed/performed image reconstruction and atom tracing. Z. L. and J. Z. analyzed the data and interpreted the results. Z. L., Y. Z., C. O. and J. Z. discussed/analyzed the 3D strain analysis. Z. L., X. M. and J. Z. wrote the manuscript. All authors commented on the manuscript.

**Competing interests** The authors declare no competing interests.

**Methods**

**Sample preparation.** Pd@Pt core-shell nanoparticles were synthesized using two-step chemical reduction procedures published elsewhere[30,53-54]. Pd(II) and Pt(IV) chlorides or acetylacetonates were dissolved in oleylamine or water at concentrations listed in Supplementary Table 1. First, the Pd(II) solutions were heated at a certain temperature to yield Pd seeds with different morphologies. Different reagents were used to grow EPB, PB and TO seeds. For EPB seeds, tri-n-octyl phosphine (TOP) was added for seeds growth. Tetrabutylammonium bromide (TBAB) and oleic acid were used for the growth of PB seeds. Poly(vinyl-pyrrolidone) (PVP), L-ascorbic acid, KBr and KCl were used for TO seeds. The resulting seeds were washed with ethanol and re-dispersed in oleylamine by ultrasonication (Supplementary Fig. 1). Second, Pt(IV) solutions were added into pre-heated (180 °C) oleylamine solutions containing different seeds to grow the Pt shell. The solutions were aged at 180 °C for 1 hrs. The final resulting CS-NPs were dispersed in cyclohexane. After being deposited onto 7-nm-thick silicon nitride membranes, the nanoparticles were baked at 180 °C for 24 hrs in vacuum to eliminate any hydrocarbon contamination. EDX analysis (Supplementary Fig. 1g) show that the nanoparticles were well defined core-shell structures with uniform size.

**Data acquisition.** ADF-STEM image series with a tilt range of ±76° for three nanoparticles were acquired on an aberration-corrected FEI Titan Themis G2 300 microscope operated at 300 kV. To minimize sample drift, three to six sequential images were obtained with a dwell time of 2 μs in each tilt angle (Supplementary Figs. 2-4). Images taken at 0° tilt angle before, during and after the acquisition of each tilt series ensured that no noticeable structural change was observed for three nanoparticles (Supplementary Fig. 5). The total electron dose of each tilt series was estimated to be between $5.6 \times 10^5$ electrons Å$^{-2}$ and $6.4 \times 10^5$ electrons Å$^{-2}$ (Supplementary Table 2).

**Image pre-processing and 3D reconstruction.** The three to six images were registered using normalized cross-correlation and then averaged at each tilt angle. Linear stage drift was estimated and corrected during the image registration. The experimental images were denoised using the block-matching and 3D filtering (BM3D) algorithm[55]. The BM3D parameters were optimized by minimizing the R factor between simulated ADF-STEM images and denoised experimental images with various denoising level. For each denoised image, a 2D mask slightly larger than the boundary of the nanoparticles was generated from each experimental image. The background level within the mask was estimated using Laplacian interpolation. Then the images in each tilt series were aligned by the center of mass and common line method.

After image pre-processing, each experimental tilt series was reconstructed by the Real Space Iterative Reconstruction (RESIRE) algorithm[28]. Angular refinement and spatial re-alignment were applied to reduce the angular errors due to sample holder rotation and stage instability. After no further improvement could be made, we performed the final reconstruction of each tilt series using RESIRE with the parameters shown in Supplementary Table 2.

**3D atomic coordinates determination and chemical species classification.** 3D atomic coordinates of three nanoparticles were determined separately using the following procedure. All local maxima in the 3D reconstruction were identified and the peak positions were located using polynomial fitting described elsewhere[56]. A minimum distance constraint of 2.2 Å was used to obtained a list of potential atoms. By applying the 3D polynomial fitting to all potential atom positions, we obtained accuracy atom positions. These positions were further manually checked to correct for unidentified or misidentified atoms due to fitting failure or a large chunk of connected intensity blobs from multiple atoms[57].

Using a K-mean clustering method[58] based on the integrated intensity of a 2.4× 2.4 × 2.4 Å$^3$ (7×7×7 voxels) volume centered at each potential atom position, all potential atoms were classified into Pd and Pt. Owing to the missing wedge and experimental noise, some surface Pt atoms with relative weak intensity were classified into Pd during K-mean clustering. To further improve the atom classification accuracy, we performed local surface re-classification and manually checked[19] a small percentage of the inconsistent atoms. The final classification results are shown in Supplementary Table 2.

The structure of two five-fold twinned nanoparticles is analyzed by two methods, i.e., local bond orientational order (BOO) parameters and polyhedral template matching[33]. Averaged local BOO parameters ($Q_4$ and $Q_6$) and normalized BOO parameter for ideal face-centered cubic (fcc) structure are calculated based on the procedure published elsewhere[29], using the first-nearest-neighbor shell distance (3.35 Å) as a constraint. By applying polyhedral template matching , PB/EPB particle was separated into

four

five fcc grain and a five-fold coaxial twin boundary which locally possessing hexagonally close-packed environment.

Supplementary Information

# Probing the atomically diffuse interfaces in core-shell nanoparticles in three dimensions

Zezhou Li[1], Zhiheng Xie[1], Yao Zhang[1], Xilong Mu[1], Haijing Yin[1], Ya-wen Zhang[1], Colin Ophus[2], Jihan Zhou[1]

[1]*Beijing National Laboratory for Molecular Sciences, College of Chemistry and Molecular Engineering, Peking University, Beijing, 100871, China.*

[2]*National Center for Electron Microscopy, Molecular Foundry, Lawrence Berkeley National Laboratory, Berkeley, CA 94720, USA.*

*Correspondence and requests for materials should be addressed to J. Z. (email: jhzhou@pku.edu.cn)*

This PDF file includes Supplementary Table 1-2 and Supplementary Figs. 1-12.

**Supplementary Table 1** | Synthesis of Pd@Pt core-shell nanoparticles

|     | Pd | | | Pt | | |
| --- | --- | --- | --- | --- | --- | --- |
|     | Precursor | Solvent | Concentration (μmol/ml) | Precursor | Solvent | Concentration (μmol/ml) |
| **PB** | PdAc$_2$ | oleylamine | 1.5 | Pt(acac)$_2$ | oleylamine | 3.3 |
| **EPB** | Pd(acac)$_2$ | oleylamine | 2.5 | H$_2$PtCl$_6$ | oleylamine | 5.6 |
| **TO** | Na$_2$PdCl$_4$ | DI water | 1.0 | H$_2$PtCl$_6$ | oleylamine | 1.2 |

**Supplementary Table 2** | Data collection, processing, reconstruction, refinement and statistics

|     | PB | EPB | TO |
| --- | --- | --- | --- |
| **Data Collection and Processing** | | | |
| Voltage (kV) | 300 | 300 | 300 |
| Convergence semi-angle (mrad) | 30.0 | 30.0 | 30.0 |
| Probe size (Å) | 0.8 | 0.8 | 0.8 |
| Detector inner angle (mrad) | 39.4 | 39.4 | 39.4 |
| Detector outer angle (mrad) | 200 | 200 | 200 |
| Pixel size (Å) | 0.343 | 0.343 | 0.343 |
| Number of projections | 67 | 59 | 60 |
| Tilt range (°) | -76.0 76.0 | -76.0 77.0 | -75.5 77.5 |
| Electron dose ($10^5$ e$^-$/Å$^2$) | 6.4 | 5.6 | 5.7 |
| **Reconstruction** | | | |
| Algorithm | RESIRE | RESIRE | RESIRE |
| Oversampling ratio | 4 | 4 | 4 |
| Number of iterations | 200 | 200 | 200 |
| **Refinement** | | | |
| R (%)[a] | 4.30 | 5.66 | 5.35 |
| **Statistics** | | | |
| # of atoms | | | |
| Total | 12038 | 5377 | 8425 |
| Pd | 2054 | 1372 | 3649 |
| Pt | 9984 | 4005 | 4776 |

[a] The R-factor is defined by $R = \frac{1}{N}\sum_\theta \frac{\sum_{x,y}|\Pi_\theta(O)\{x,y\}-b_\theta\{x,y\}|}{\sum_{x,y}|b_\theta\{x,y\}|}$, where $\Pi_\theta(O)\{x, y\}$ is the back projection of the reconstruction volume at angle $\theta$, $b_\theta\{x, y\}$ is the real projection image at angle $\theta$, and $N$ is the number of projections.

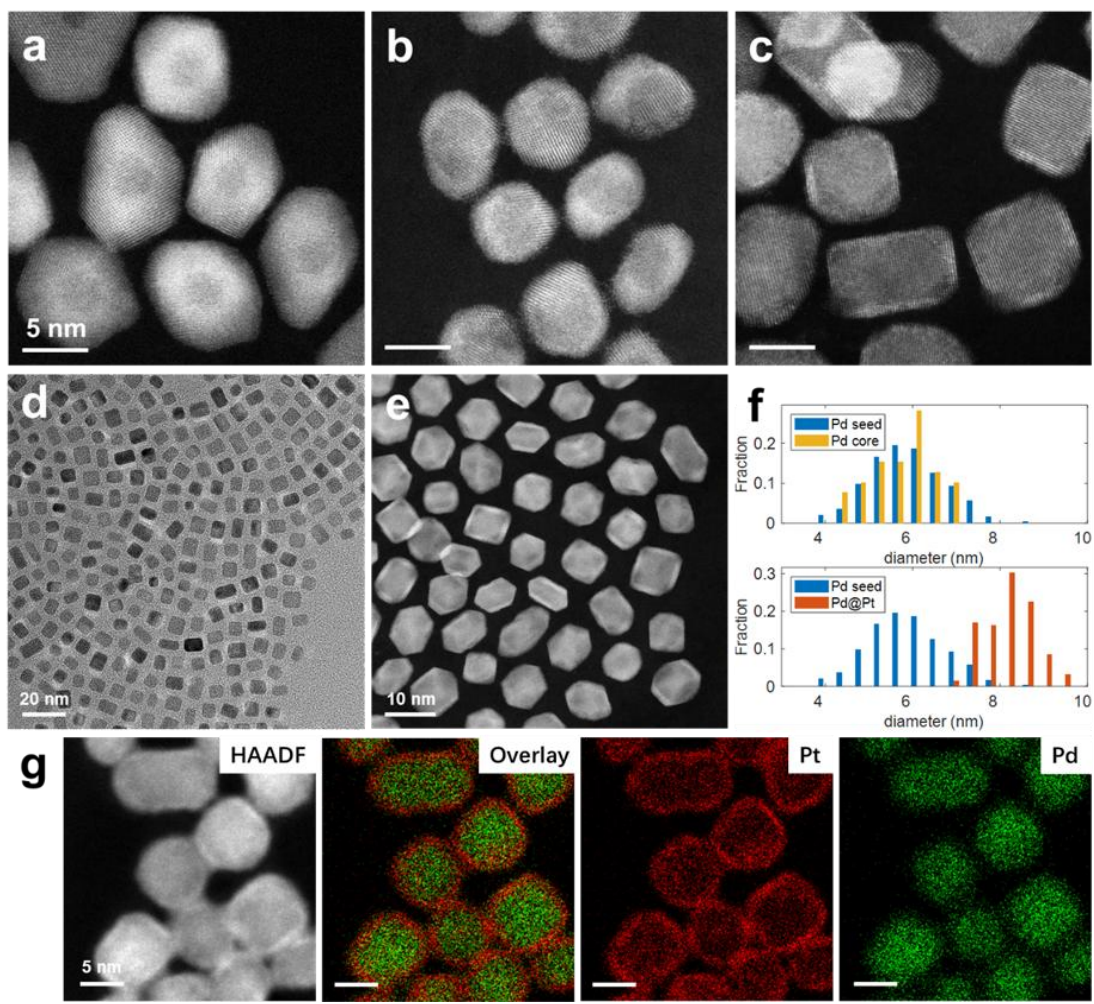

**Supplementary Fig. 1 | Characterization of Pd@Pt core-shell nanoparticles synthesized using a two-step chemical reduction procedure. a-c**, HAADF-STEM images of PB (**a**), EPB (**b**) and TO (**c**) particles, respectively. **d**, Large scale TEM image of cuboid Pd seed. **e**, Large scale HAADF-STEM image of TO structured Pd@Pt CS-NPs synthesized with Pd seeds in **d**. **f**, Particle size statistics: while the size of Pd seeds is significantly smaller than that of Pd@Pt CS-NPs, the size distribution of both the Pd seeds and the Pd core in Pd@Pt CS-NPs are almost identical. **g**, EDX mapping of Pd@Pt CS-NPs showing well-defined core-shell structure at nanometer scale.

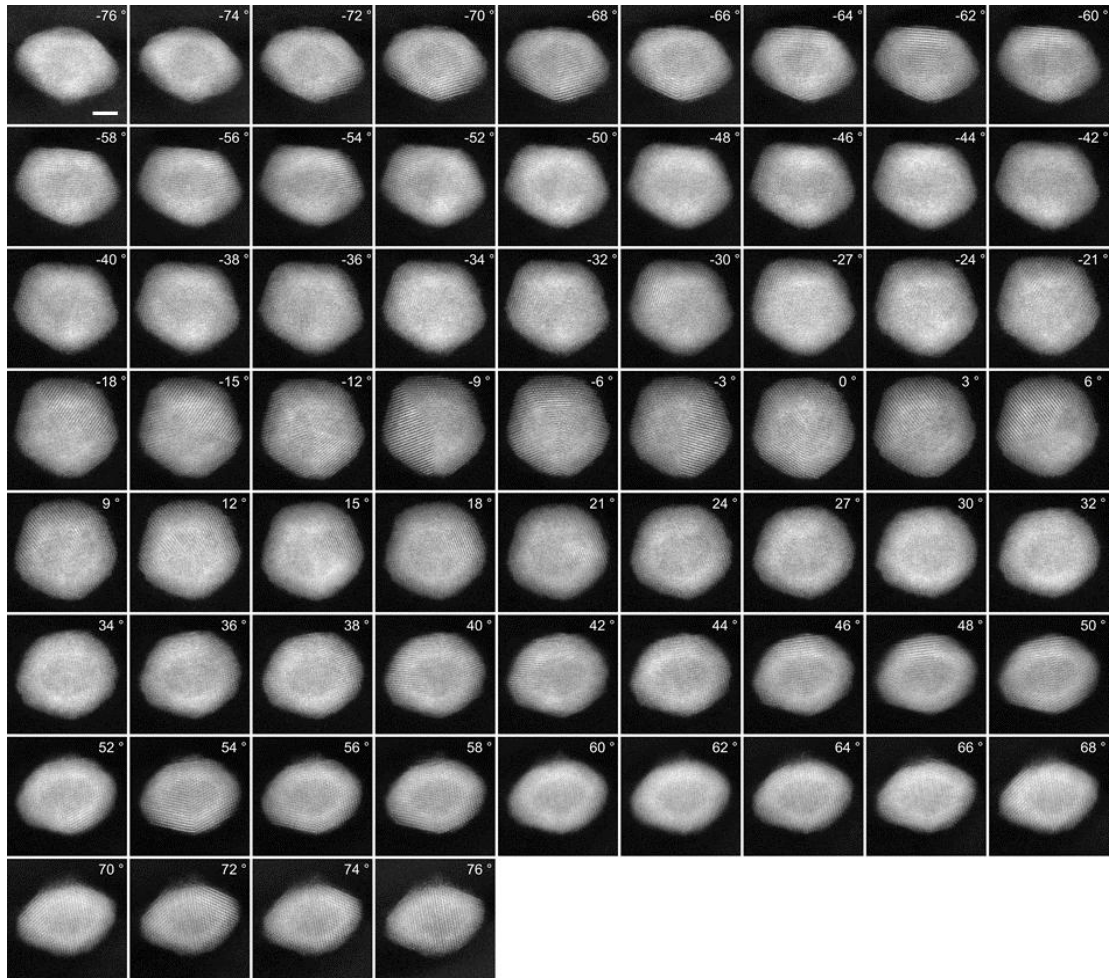

**Supplementary Fig. 2 | Tomographic tilt series of PB particle.** 67 ADF-STEM images with a tilt range from −76.0° to +76.0°. Scale bar, 2 nm.

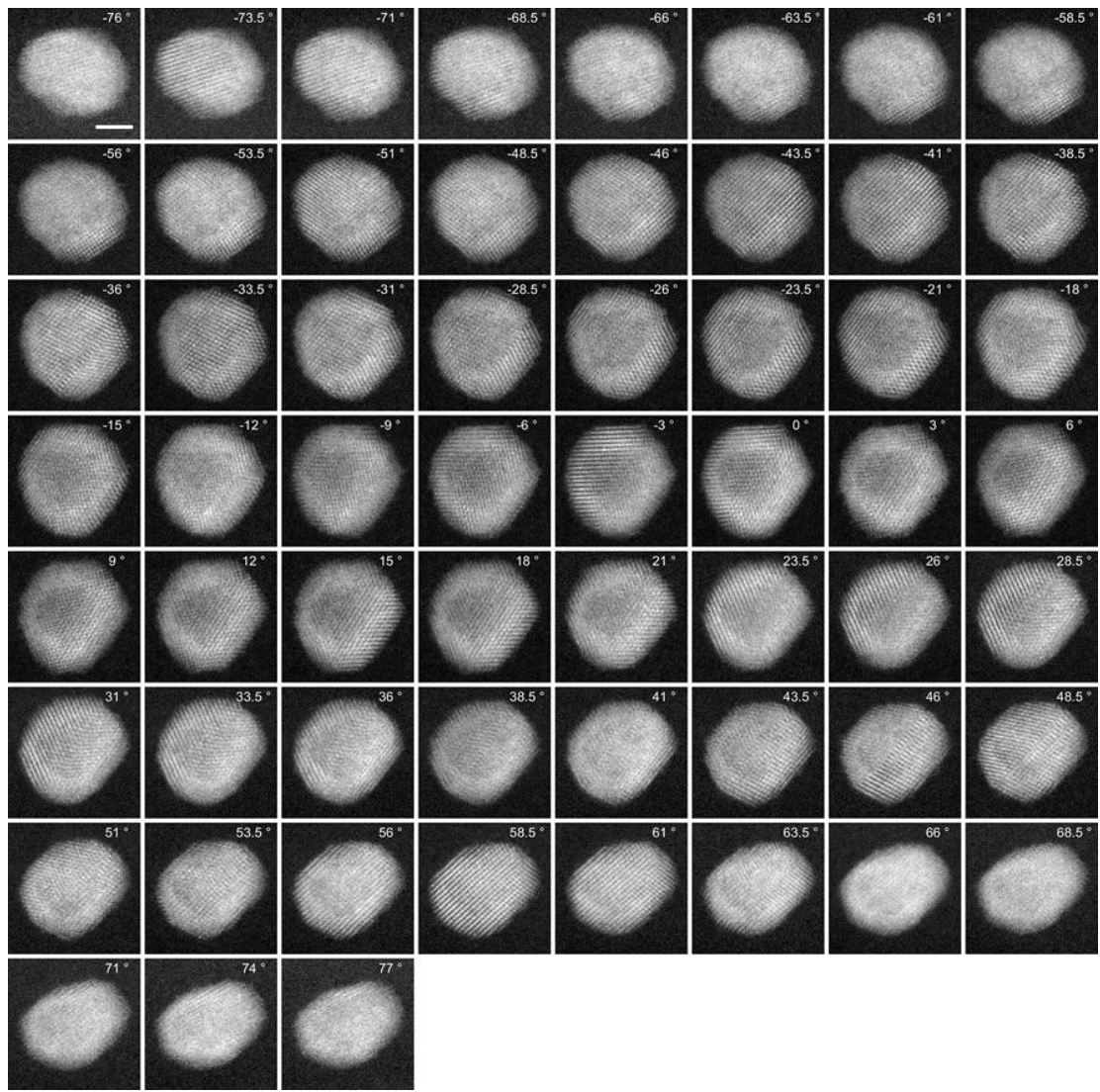

**Supplementary Fig. 3 | Tomographic tilt series of EPB particle.** 59 ADF-STEM images with a tilt range from −76.0° to +77.0°. Scale bar, 2 nm.

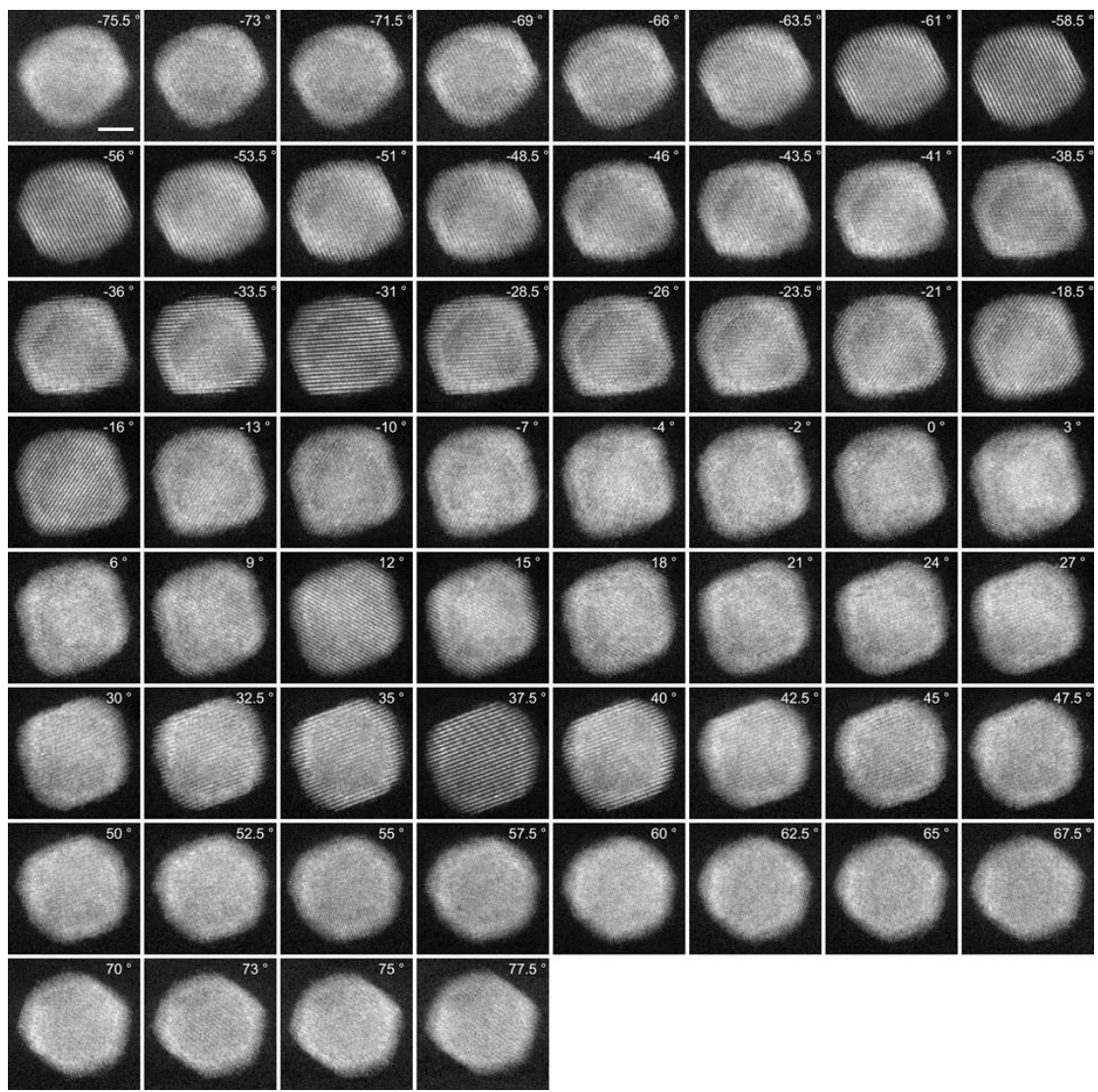

**Supplementary Fig. 4 | Tomographic tilt series of TO particle.** 60 ADF-STEM images with a tilt range from −75.5° to +77.5°. Scale bar, 2 nm.

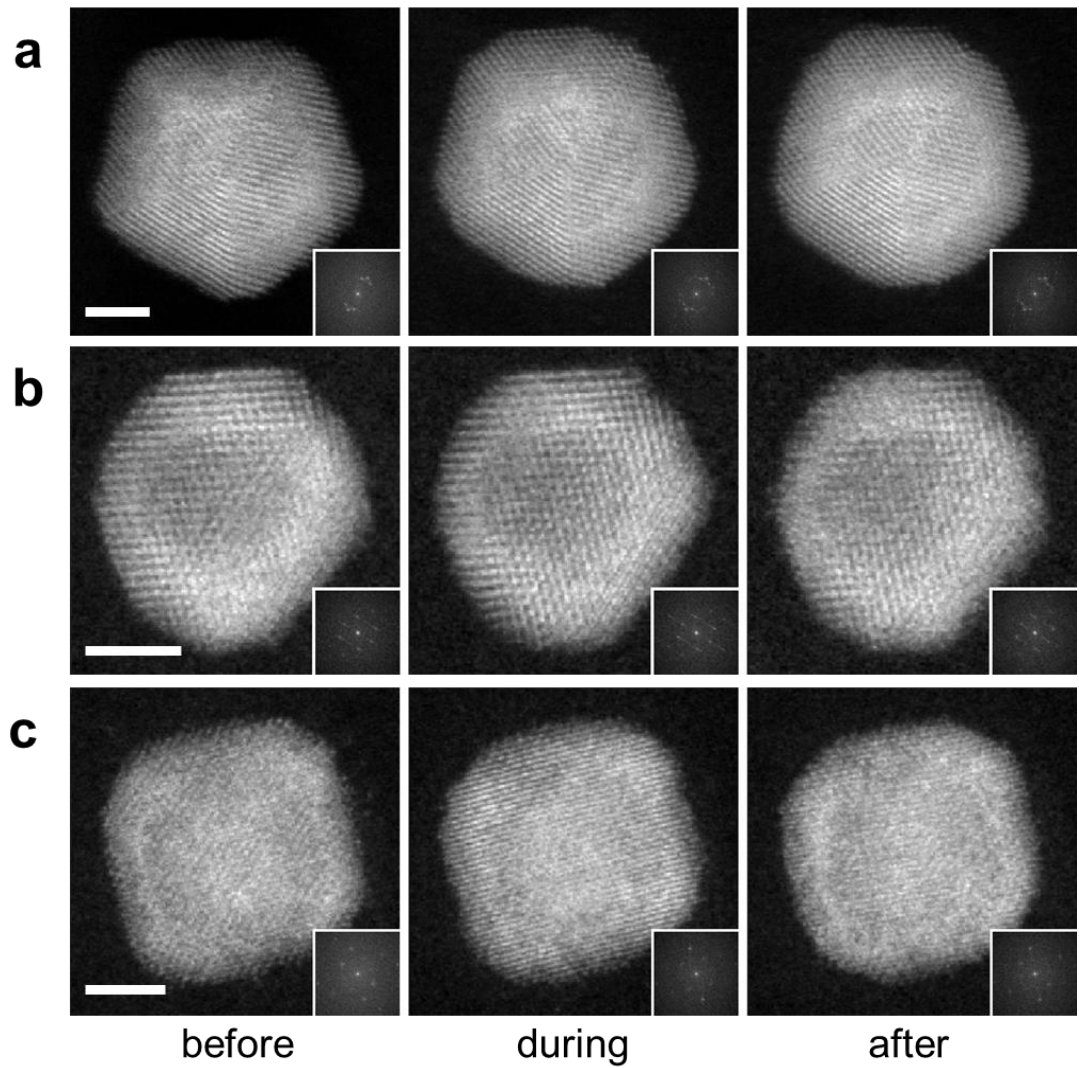

**Supplementary Fig. 5 | Consistency check of three nanoparticles. a-c**, ADF-STEM images taken at 0° before, during and after tilting experiment for PB (**a**), EPB (**b**) and TO (**c**) nanoparticles. Scale bar, 2 nm.

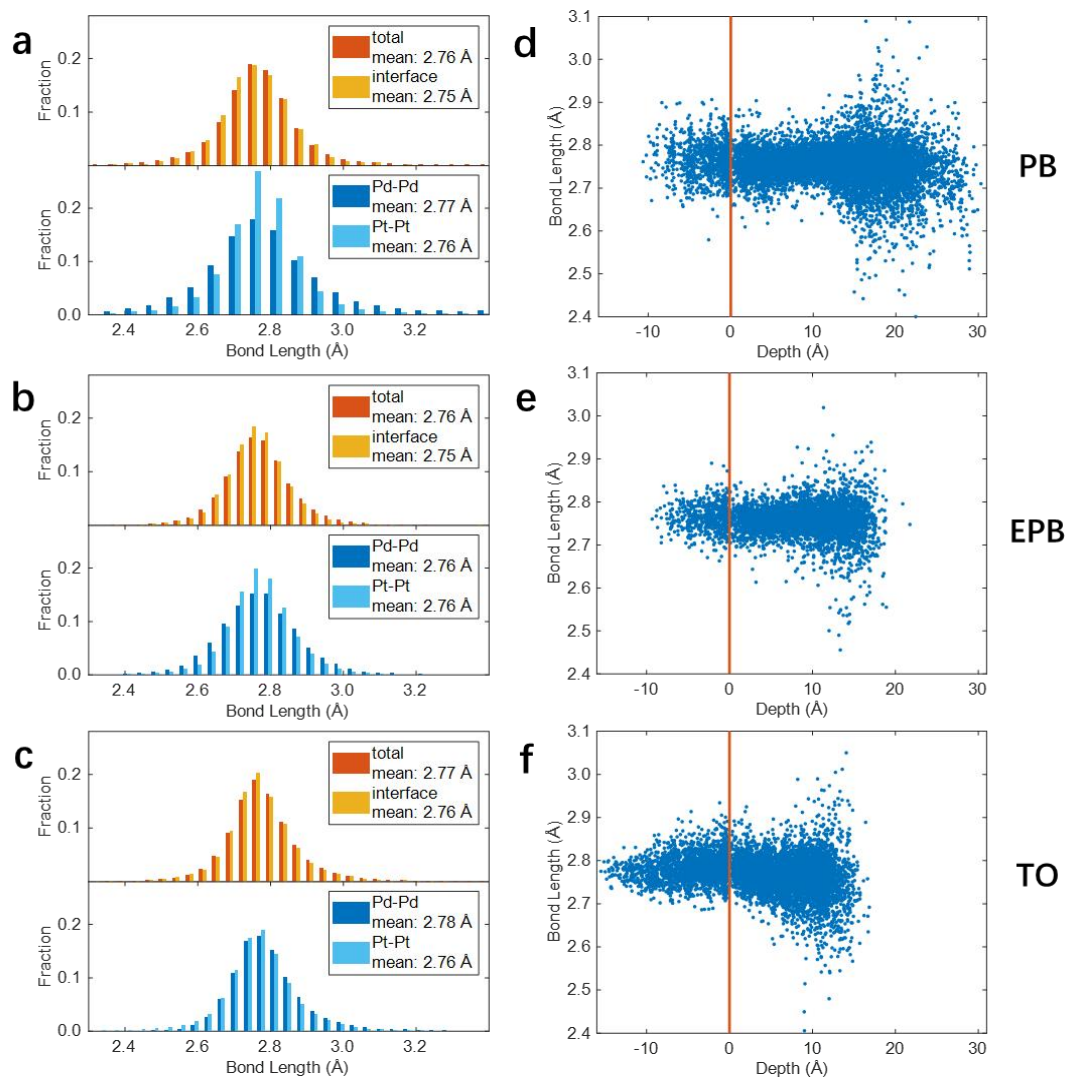

**Supplementary Fig. 6 | Bond length statistics of three particles. a-c**, Total and interfacial Pd-Pd and Pt-Pt bond length distribution in PB (**a**), EPB (**b**) and TO (**c**) particles. **e-f**, Distribution of bond length versus depth in the particles. Zero depth (red line) is defined at the depth where the mean Pd concentration is ~50 % in each particle.

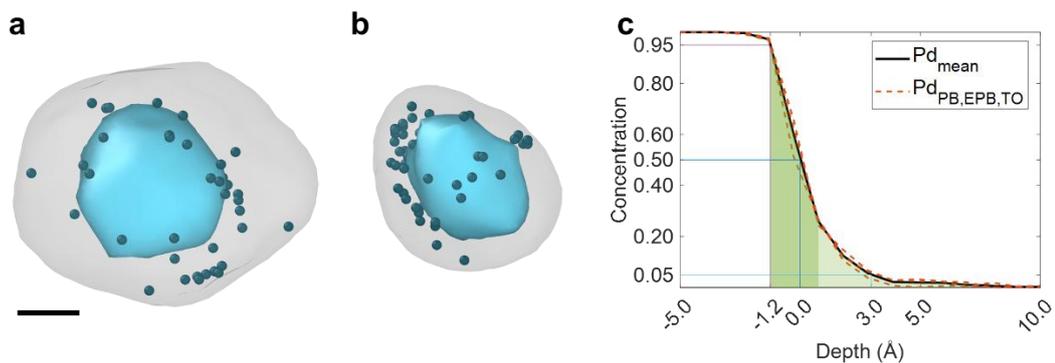

**Supplementary Fig. 7 | Concentration and distribution of Pd in three particles.**
**a-b**, Distribution of isolated Pd atoms in PB (**a**) and EPB (**b**) particles, respectively. Scale bar for **a-b**, 2 nm. **c**, Radially-averaged Pd concentrations along the core to the shell for three particles, where the diffuse interface is highlighted with green and pale green. The green represent the region where Pd concentration drops fast from the core to the shell. The pale green represent the region where Pd concentration decreases slowly. Most of isolated Pd atoms distribute in the pale green area. We measured the thickness of the CS interface using Pd concentration between 5% and 95% as the defined thickness range. Zero depth was defined as where Pd concentration equals to 50%.

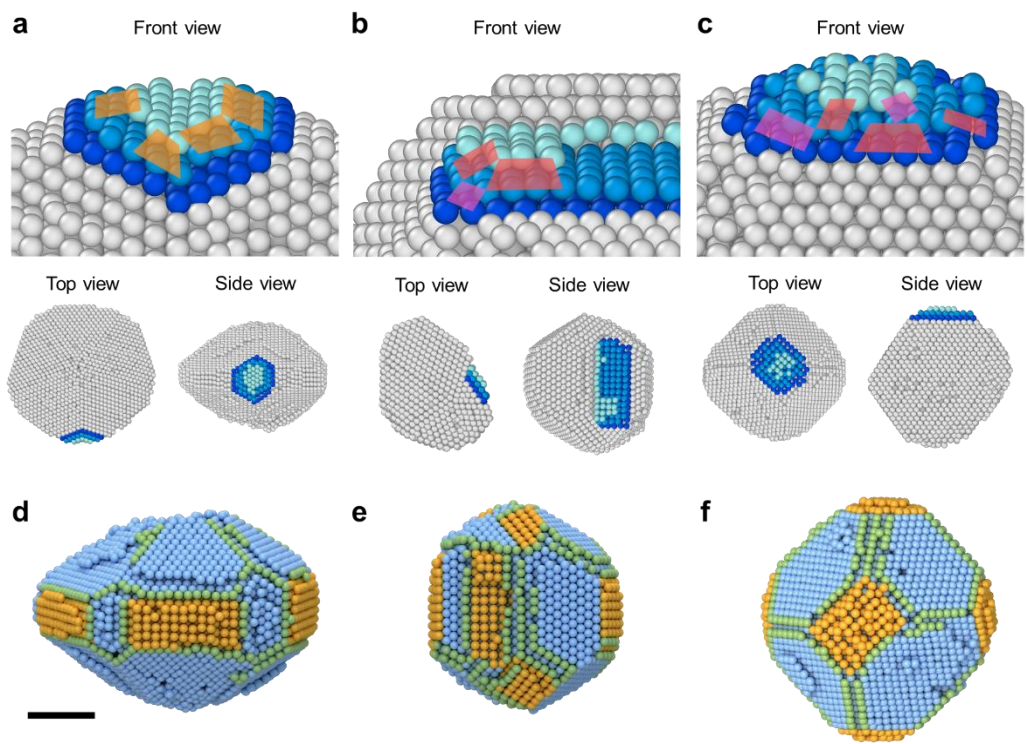

**Supplementary Fig. 8 | Ridge and edge structures in three particles. a-c**, The top panel show front views of three enlarged corners of PB (**a**), EPB (**b**) and TO (**c**) particles, respectively. The first, second and third layer atoms are colored in pale cyan, light blue and dark blue, respectively. The bottom panel show the top view and side view of three whole particles. Two connected {111} facets are colored in PB (**a**). One {100} facet is colored in EPB (**b**) and TO (**c**), respectively. Orange, magenta and red patches represent (S)-[2(111) × (110)], (S)-[2(100) × (100)], (S)-[2(100) × (110)], respectively. **d-f**, 3D atomic models of EPB (**d**), PB (**e**) and TO (**f**) particles, respectively. The {100} and {111} facets atoms are colored in yellow and blue, respectively. The atoms of the ridges and edges on the surface are colored in green. Scale bar for **d-f**, 2nm.

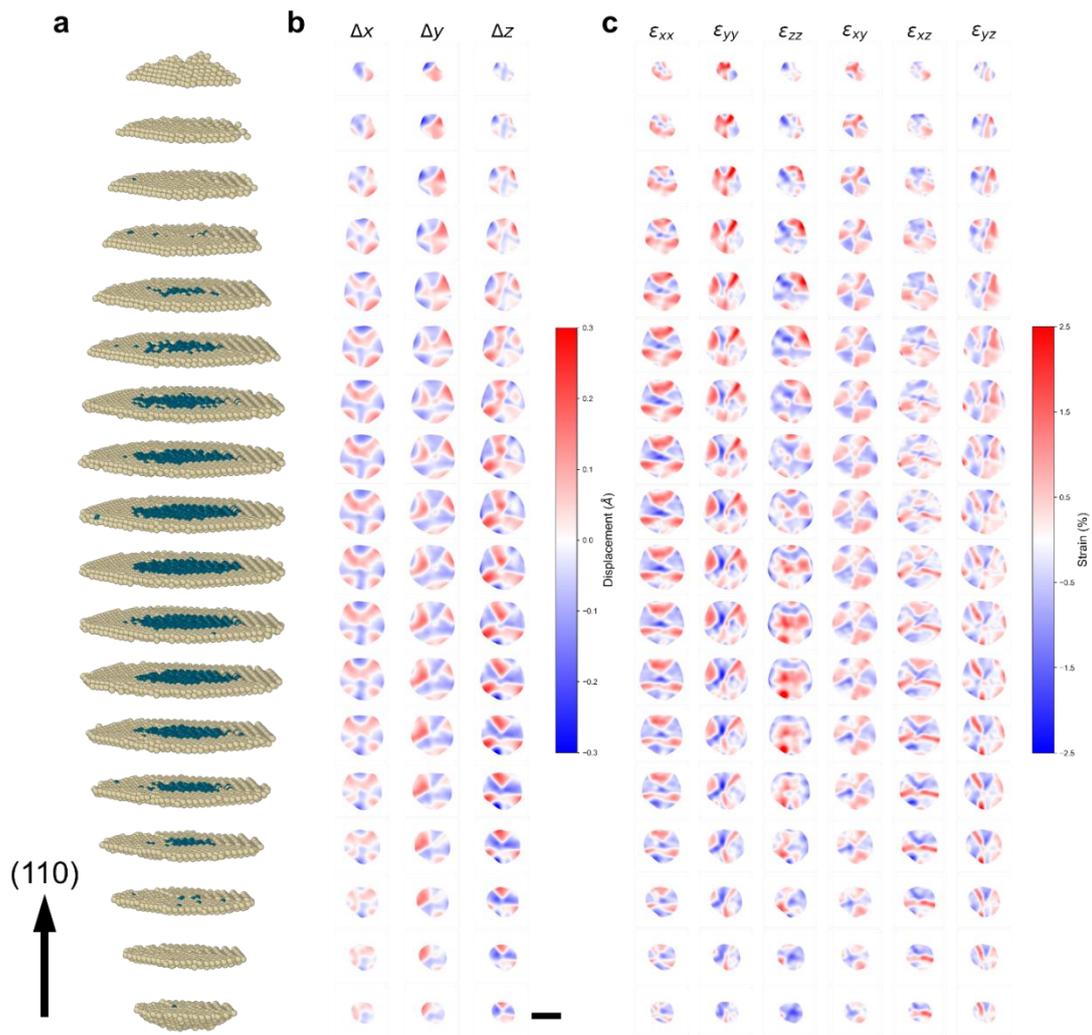

**Supplementary Fig. 9 | 3D atomic displacements and strain maps of particle PB.** Atomic slices (**a**), 3D displacement field (**b**) and six components of the full strain tensor (**c**) of PB particle. Scale bar for **b** and **c**, 5 nm.

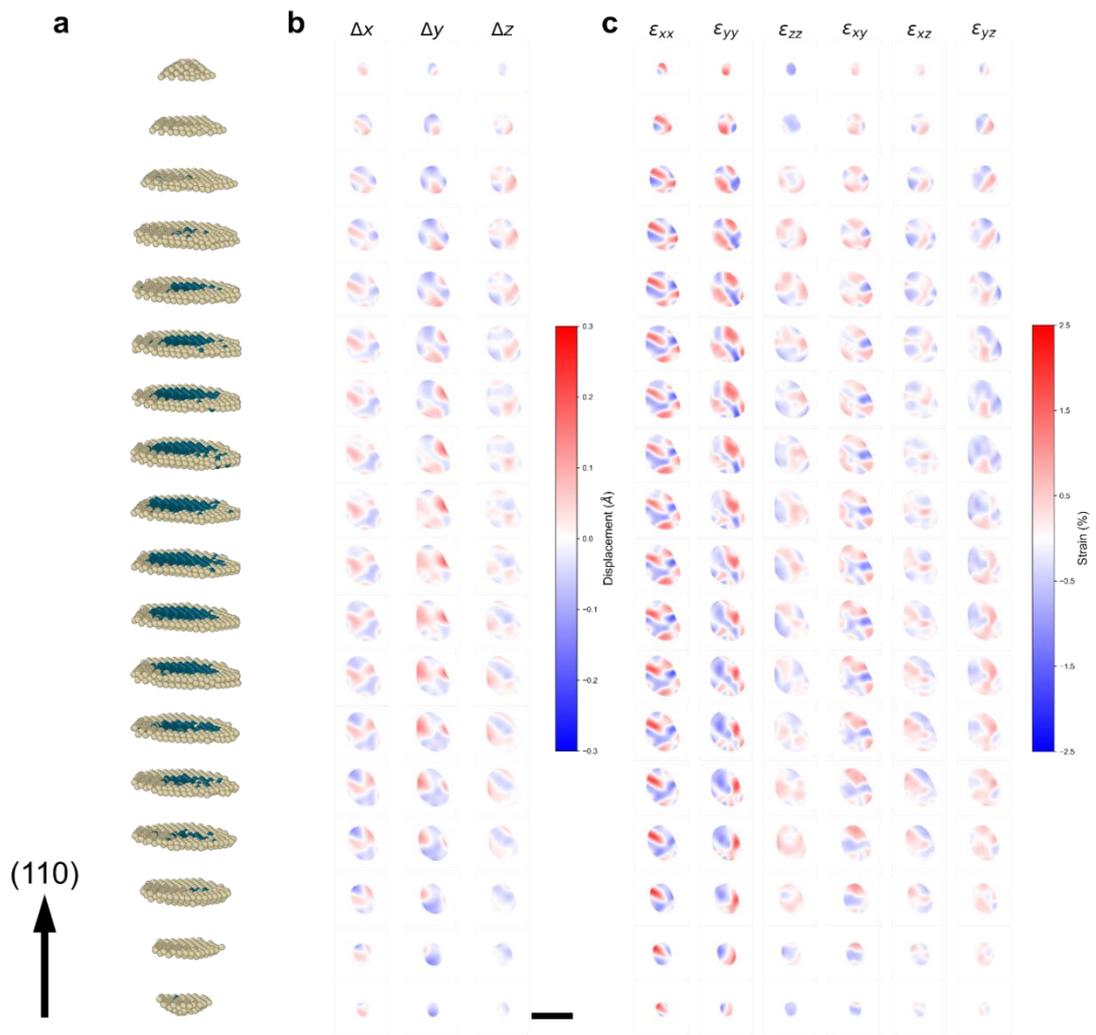

**Supplementary Fig. 10 | 3D atomic displacements and strain maps of particle EPB.** Atomic slices (**a**), 3D displacement field (**b**) and six components of the full strain tensor (**c**) of EPB particle. Scale bar for **b** and **c**, 5 nm.

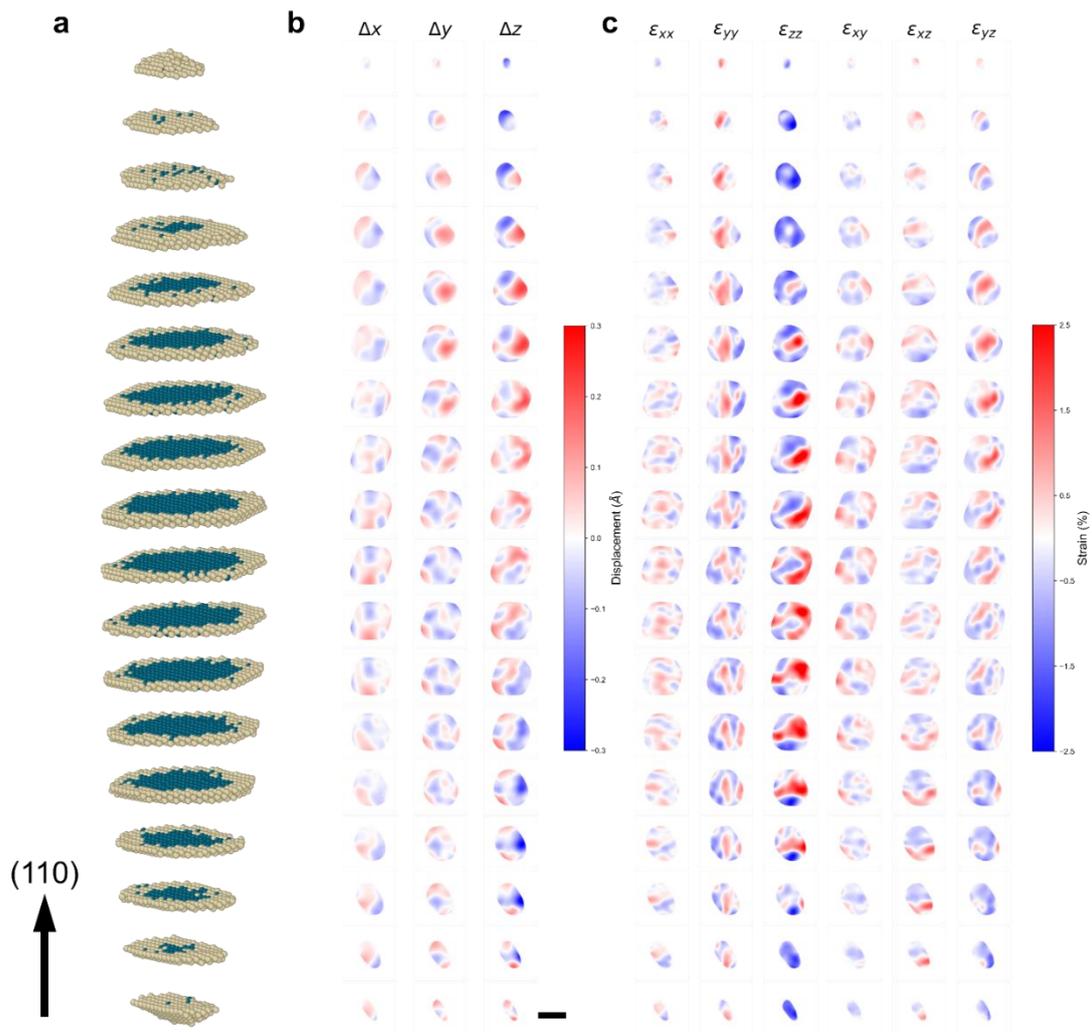

**Supplementary Fig. 11 | 3D atomic displacements and strain maps of particle TO.** Atomic slices (**a**), 3D displacement field (**b**) and six components of the full strain tensor (**c**) of TO particle. Scale bar for **b** and **c**, 5 nm.

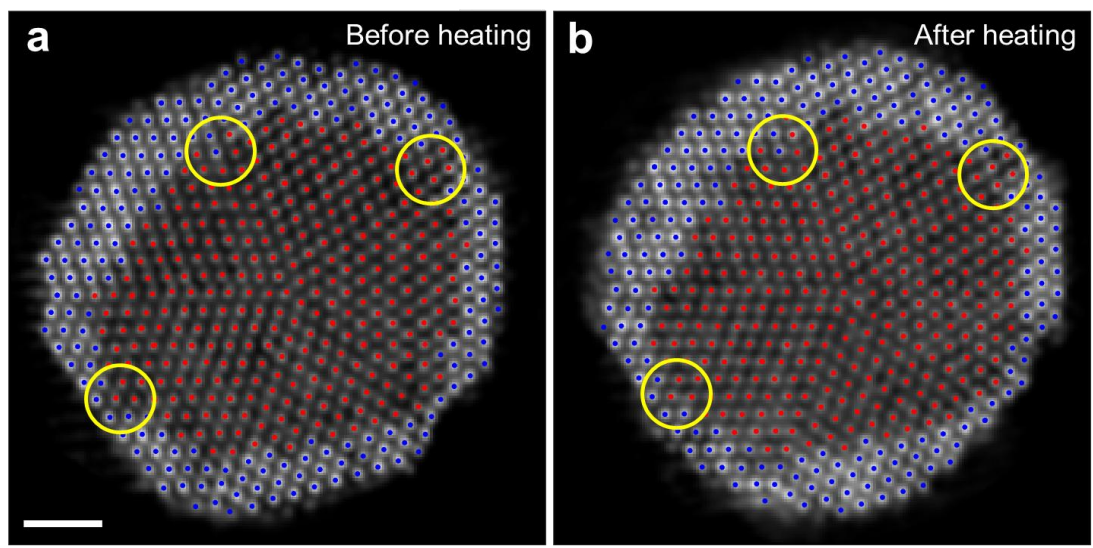

**Supplementary Fig. 12 | The influence of long times heating procedure on the atomically diffuse interfaces.** The same atomic layer of a Pd@Pt CS-NP before (**a**) and after (**b**) heating in the vacuum at 180 °C for 48 hrs, obtained from the two independent measurements. Yellow circles highlight consistent diffuse interface. Scale bar for **a** and **b**, 1 nm. The two independent tomographic tilt series were acquired from the same nanoparticle before and after long term baking at 180 °C. The 3D atomic models from the two measurements were obtained using the same reconstruction, atom tracing, atom identification and refinement procedures. Although some of the surface atoms are inconsistent in the two models due to the surface atom rearrangement during heating and experimental error, the internal atoms belonging to the interface are almost identical.